\documentclass[twocolumn,amsmath,showpacs,amsfonts,aps,prc,floatfix]{revtex4}
\usepackage{graphicx}
\usepackage{bm}

\newcommand{\la}{\langle}
\newcommand{\ra}{\rangle}
\begin{document}

\title{ Even-by-event hydrodynamical simulations for $\sqrt{s}_{NN}$=200 GeV Au+Au collisions and the correlation between flow coefficients and initial asymmetry measures}

\author{A. K. Chaudhuri}
\email[E-mail:]{akc@vecc.gov.in}
\affiliation{Theoretical Physics Division, Variable Energy Cyclotron Centre,
1/AF, Bidhan Nagar, 
Kolkata 700~064, India}

 \author{Md. Rihan Haque}
\email[E-mail:]{rihanphys@vecc.gov.in}
\affiliation{Variable Energy Cyclotron Centre, 1/AF, Bidhan Nagar, 
Kolkata 700~064, India}

\author{Victor Roy}
\email[E-mail:]{victor.physics.pm@gmail.com}
\affiliation{School of Physical Sciences, National Institute of Science Education and Research, Bhubaneswar-751005, India}

\author{Bedangadas Mohanty}
\email[E-mail:]{bedanga@rcf.rhic.bnl.gov}
\affiliation{School of Physical Sciences, National Institute of Science Education and Research, Bhubaneswar-751005, India}
 
\begin{abstract}
Centrality dependence of charged particles multiplicity, transverse momentum spectra, integrated and  differential elliptic flow,  in $\sqrt{s}_{NN}$=200 GeV Au+Au collisions 
 are analyzed using event by event ideal hydrodynamics. Monte-Carlo Glauber model of initial condition, constrained to reproduce experimental charged particle's multiplicity in 0-10\% Au+Au collisions, reasonably well reproduces all the experimental observables, e.g. centrality dependence of charged particles multiplicity, integrated and differential elliptic flow.  Model predictions for higher flow harmonics, $v_3$, $v_4$ however overestimate the experimental data, more in the peripheral collisions than in the central collisions. Correlation between initial (spatial) asymmetry measures and flow coefficients are also studied. With exception of the elliptic flow, for all the higher flow coefficients ($v_n$,n=3-5), correlation is reduced with collision centrality. In peripheral collisions, higher flow coefficients are only weakly correlated to the asymmetry measures. Elliptic flow however, remains strongly correlated with initial eccentricity in all the collision centralities.
\end{abstract}

\pacs{47.75.+f, 25.75.-q, 25.75.Ld} 

\date{\today}  

\maketitle

\section{Introduction}\label{sec1}
 
It is expected that collisions between two nuclei at
ultra-relativistic energies will lead to a phase transition
from hadrons to the fundamental constituents, quarks
and gluons, usually referred to as Quark-Gluon-Plasma
(QGP). Experiments at the Relativistic Heavy Ion Collider (RHIC) at $\sqrt{s}_{NN}$=200 GeV Au+Au collisions \cite{BRAHMSwhitepaper} \cite{PHOBOSwhitepaper}\cite{PHENIXwhitepaper} \cite{STARwhitepaper} and at the Large Hadron Collider (LHC) at $\sqrt{s}_{NN}$=2.76 TeV Pb+Pb collisions \cite{Aamodt:2010pb}\cite{Collaboration:2010cz}\cite{Aamodt:2010jd}\cite{Aamodt:2010pa} had provided compelling evidences for production of QGP.
One of the experimental observables of QGP is
the azimuthal distribution of produced particles. 
 It is best studied by decomposing it in a Fourier series, 
 
\begin{equation} \label{eq1}
\frac{dN}{d\phi}=\frac{N}{2\pi}\left [1+ 2\sum_n v_n cos(n\phi-n\psi)\right ], n=1,2,3...
\end{equation} 
 
\noindent   $\phi$ is the azimuthal angle of the detected particle and 
$\psi$ is the  plane of the symmetry of initial collision zone. In $\sqrt{s}_{NN}$=200 GeV Au+Au collisions,   second flow harmonic ($v_2$), usually referred to as the elliptic flow, has been extensively studied experimentally as well as theoretically. Experimentally observed finite, non-zero $v_2$ is now regarded as definite proof of collective QCD matter creation in Au+Au collisions.
Qualitatively, elliptic
flow is naturally explained in a hydrodynamical model,
rescattering of secondaries generates pressure and drives
the subsequent collective motion. In non-central collisions, the reaction zone is asymmetric (almond shaped),
pressure gradient is large in one direction and small in the
other. The asymmetric pressure gradient generates the
elliptic flow. As the fluid evolve and expands, asymmetry in the reaction zone decreases and a stage arises when the
reaction zone become symmetric and system no longer
generates elliptic flow. Elliptic flow is early time phenomena. It is a sensitive probe to, (i) degree of thermalisation,
(ii) transport coefficient and (iii) equation of state of the
early stage of the fluid.

Ideal and viscous hydrodynamic models have been extensively used to analyze the experimental data at RHIC and LHC energy collisions.
Most of the analyses were performed with smooth
initial matter distribution   obtained from geometric overlap of density distributions of the colliding nuclei.
For smooth matter distribution, the plane of symmetry of the collision zone coincides with the reaction plane (the plane containing the impact parameter and the beam axis).
The odd Fourier coefficients are zero by symmetry. One of the important realization in recent years, is that the participating nucleons, rather than the reaction plane,
determines the symmetry plane of the initial collision zone \cite{Manly:2005zy}. The realization is the results of analysis of various experimental data, e.g. the two particle correlation in $\Delta \phi$-$\Delta \eta$ plane \cite{Adams:2005ph}\cite{Putschke:2007mi}\cite{Abelev:2008ac}. The peculiar structure in two particle correlations known as 'ridge' and 'shoulder', observed both in STAR and PHENIX experiments have most compelling explanation provided the third flow harmonic, the triangular flow $v_3$ develops in the collisions. 
Specifically, if initial condition is parameterized with quadrapole and triangular moments,  response of the medium to these anisotropies is reflected in the two body correlation as ridge and shoulder \cite{Alver:2010gr},\cite{Alver:2010dn}.
 Importance of the higher order flow harmonics in explaining the peculiar structures in two body correlation was also argued by  Sorensen \cite{Sorensen:2010zq}.
 The ridge structure in $p{\bar p}$ collisions  \cite{Khachatryan:2010gv} \cite{Velicanu:2011dp} also has a natural explanation if odd harmonic flows develop.  Recently, ALICE collaboration has observed odd harmonic flows    in Pb+Pb collisions \cite{:2011vk}. In most central collisions, the elliptic flow ($v_2$) and triangular flow ($v_3$) are of similar magnitude. In peripheral collisions however, elliptic flow dominates. More recently, PHENIX collaboration  \cite{Adare:2010ux}\cite{Adare:2011tg}\cite{Lacey:2011av} measured triangular flow in $\sqrt{s}_{NN}$=200 GeV Au+Au collisions.

In the present paper, in event-by-event hydrodynamics, with Monte-Carlo Glauber model initial energy density distribution, we have simulated $\sqrt{s}_{NN}$=200 GeV Au+Au collisions in 0-10\% to 40-50\% collision centralities. 
 Simulation results compare well with the existing experimental data on charged particles multiplicity, transverse momentum spectra at $p_T\leq 1 GeV$, integrated and differential elliptic flow. Higher flow coefficients however are over predicted. 
We have also studied the centrality dependence of 
  the correlation between the (integrated) flow coefficients with the initial spatial asymmetry measures. 
  In 0-10\%-40-50\% collisions, elliptic flow remains strongly correlated with the initial eccentricity. Triangular flow is strongly correlated with initial triangularity only in very central collisions. The correlation is reduced significantly in peripheral collisions. Higher flow coefficients $v_n$, n=4-5, even in central collisions, is only weakly correlated with initial asymmetry measures, $\epsilon_n$, n=3-5 and the  correlation is more reduced in peripheral collisions.
  
  The paper is organized as follows; in section \ref{sec2}, the Monte-Carlo Glauber model for initial energy density for use in hydrodynamic simulations is briefly discussed. In section \ref{sec3}, hydrodynamic equations, initial conditions, equation of state used in the simulations are described. Results of the simulations are described in section \ref{sec4}. Finally,   all the results are summarized in section \ref{sec5}.

\section{Monte-Carlo Glauber model of initial energy density distribution}\label{sec2}

In theoretical simulations of event-by-event hydrodynamics, one generally uses the Monte-Carlo Glauber model to obtain the initial energy density distribution in an event. Details of the Monte-Carlo Glauber model can be found in \cite{Alver:2008aq}.
In a Monte-Carlo Glauber model, according to the density distribution of the colliding nuclei,    two nucleons are randomly chosen. If the transverse separation  
between the two nucleons is below $\sqrt\frac{\sigma_{NN}}{\pi}$, they are assumed to interact. Where $\sigma_{NN}$ is the nucleon-nucleon interaction cross section, taken here as 42 mb for Au-Au collisions at $\sqrt{s_{NN}}$ = 200 GeV.
  Transverse position of the participating nucleons is then known in each event. The positions will fluctuate from event-to-event. If a particular event has $N_{part}$ participants,  participants positions in the transverse plane can be labeled as, $(x_1,y_1), (x_2,y_2)....(x_{N_{part}},y_{N_{part}})$. Energy density distribution in the particular event can be obtained by assuming that    each participant deposit energy $\varepsilon_0$ in the transverse plane,  

\begin{equation}\label{eq2}
\varepsilon(x,y) \approx \varepsilon_0 \sum_{i=1}^{N_{part}}  \delta(x-x_i,y-y_i)
\end{equation}

Fluid dynamical models require continuous density distribution and discrete distribution as in Eq.\ref{eq2} cannot be evolved in a hydrodynamical model. To use in a hydrodynamic model, the discrete density distribution has to be converted into a smooth energy-density distribution. This can be done by smearing the discrete participating nucleon positions by some smoothing function, $\delta(x-x_i,y-y_i) \rightarrow g(x-x_i,y-y_i,\zeta_1,\zeta_2..)$,
$\zeta_i$ being parameters of the smoothing function $g$. 
 
\begin{equation} \label{eq3}
\varepsilon(x,y)=\varepsilon_0 \sum_{i=1}^{N_{part}}  g(x-y,x_i,y-y_i,\zeta_1,\zeta_2....)
\end{equation}

One generally uses a Gaussian smoothing function. However, there can be other choices, e.g. in \cite{RihanHaque:2012wp}, a Woods-Saxon distribution function was used for the smoothing. In the present simulations, we have used a Gaussian distribution 
 
\begin{equation}
g_{gauss}(x-x_i,y-y_i,\sigma) \propto e^{-\frac{{(x-x_i)^2+(y-y_i)^2}}{2\sigma^2}  }, \label{eq6}
\end{equation}
  
\noindent of width $\sigma$=0.5 fm.

\begin{figure}[t]
 \center
 \resizebox{0.35\textwidth}{!}{%
  \includegraphics{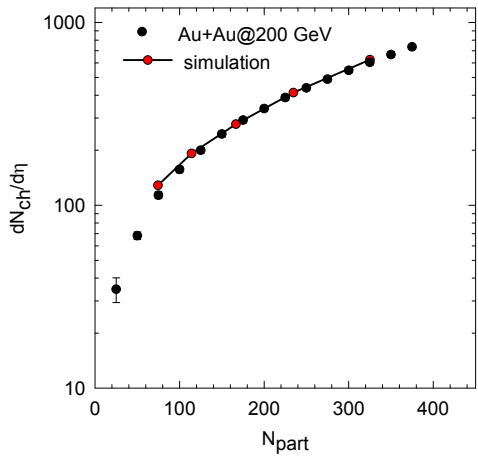} 
}
\caption{(color online) The black circles are the PHENIX data for the centrality dependence of charged particles multiplicity in $\sqrt{s}_{NN}$=200 GeV Au+Au collisions. The red symbols are the multiplicity in simulated events. The lines are to guide the eye. }
\label{F1}
\end{figure}

\section{Hydrodynamic equations, equation of state, initial conditions}\label{sec3}

With the Monte-Carlo Glauber model initial condition for
 initial energy density, space-time evolution of the fluid, in each event,  is obtained by solving the energy-momentum conservation equations,

\begin{eqnarray}\label{eq7}
T^{\mu\nu}&=&(\varepsilon+p)u^\mu u^\nu -P g^{\mu\nu}, \\
\partial_\mu T^{\mu\nu}&=&0,
\end{eqnarray}

\noindent where $\varepsilon$ and $p$ are the energy density and pressure respectively, $u$ is the hydrodynamic 4-velocity. We have assumed ideal fluid formation and disregarded any dissipative effect. 
Assuming boost-invariance, hydrodynamic equations are solved in $(\tau=\sqrt{t^2-z^2},x,y,\eta_s=\frac{1}{2}\ln\frac{t+z}{t-z})$ coordinate system, with the code AZHYDRO-KOLKATA  \cite{Chaudhuri:2008sj}. 

Hydrodynamics equations are closed with an equation of state (EoS) $p=p(\varepsilon)$.
Currently, there is consensus that the confinement-deconfinement transition is a cross over. The cross over or the pseudo critical temperature for the quark-hadron transition  is
$T_c\approx$170 MeV \cite{Aoki:2006we,Aoki:2009sc,Borsanyi:2010cj,Fodor:2010zz}.
In the present study, we use an equation of state where the Wuppertal-Budapest \cite{Aoki:2006we,Borsanyi:2010cj} 
lattice simulations for the deconfined phase is smoothly joined at $T=T_c=174$ MeV, with hadronic resonance gas EoS comprising of all the resonances below mass $m_{res}$=2.5 GeV. Details of the EoS can be found in \cite{Roy:2011xt}.

In addition to the initial energy density for which we use the   Monte-Carlo Glauber model, solution of hydrodynamic  equations   requires to specify the thermalisation or the initial time $\tau_i$ and fluid velocity ($v_x(x,y),v_y(x,y)$) at the initial time.  A freeze-out prescription is also needed to convert the information about fluid energy density and velocity to invariant particle distribution.  We assume that the fluid is thermalized at $\tau_i$=0.6 fm and the initial fluid velocity is zero, $v_x(x,y)=v_y(x,y)=0$. The freeze-out temperature is fixed at $T_F$=130 MeV. 
We use Cooper-Frye formalism to obtain the invariant particle distribution of $\pi^-$ from the freeze-out surface. Resonance production is included. Considering that pions constitute $\sim$ 20\% of all the charged particles, $\pi^-$ invariant distribution is multiplied by the factor  $2\times 1.2$ to approximate the charged particle's invariant distribution. 
From the  invariant distribution, harmonic flow coefficients are obtained as \cite{arXiv:1104.0650},

\begin{figure}[t]
\center
\resizebox{0.35\textwidth}{!}{%
\includegraphics{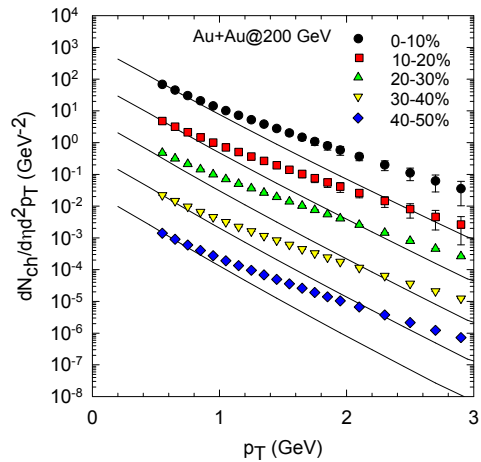} 
}
\caption{(color online) Charged particles transverse momentum distribution with and without    multiplicity fluctuations. }
\label{F2}
\end{figure}

\begin{eqnarray}
v_n(y,p_T)e^{in\psi_n(y,p_T)}&=&\frac{\int d\phi e^{in\phi} \frac{dN}{dy  p_Tdp_T d\phi}}  {\frac{dN}{dy p_Tdp_T}} \label{eq8}\\
  v_n(y)e^{in\psi_n(y)}&=& \frac{ \int p_T dp_T d\phi e^{in\phi} \frac{dN}{dy p_T dp_T d\phi} } { \frac{dN}{dy} } \label{eq9}
\end{eqnarray}
  
In a boost-invariant version of hydrodynamics, flow coefficients are rapidity independent.
Present simulations are suitable only for central rapidity, $y\approx$0, where boost-invariance is most justified. Hereafter, we drop the rapidity dependence. $\psi_n$ in Eqs.\ref{eq8},\ref{eq9} is the participant plane angle for the n-th flow harmonic. 
We characterise the asymmetry of the initial collision zone in terms of various moments of the eccentricity  \cite{Alver:2010gr},\cite{Alver:2010dn},\cite{Teaney:2010vd},

\begin{figure}[t]
\vspace{0.3cm} 
\center
\resizebox{0.35\textwidth}{!}{%
\includegraphics{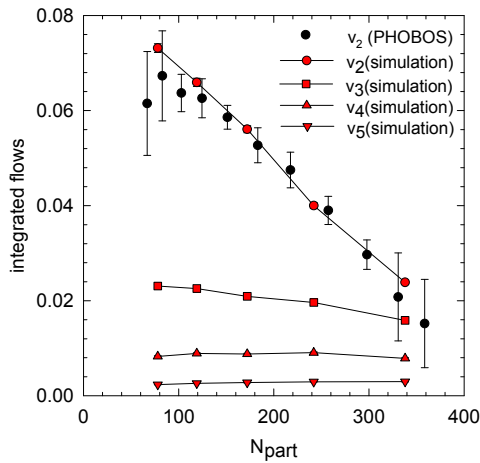}
}
\caption{(color online) Black circles are PHOBOS measurements
for the centrality dependence of elliptic flow in $\sqrt{s}_{NN}$=200 GeV Au+Au collisions. The simulation results for event averaged elliptic flow are shown as the red circles. The red squares, up triangles and down triangles are simulation results for triangular flow ($v_3$), rectangular flow ($v_4$) and pentangular flow ($v_5$) respectively.  }
\label{F3}
\end{figure} 

\begin{eqnarray} 
\epsilon_n e^{in\psi_n} &=&-\frac{\int \int \varepsilon(x,y) r^n e^{i2\phi}dxdy}{\int \int\varepsilon(x,y) r^n dxdy} \label{eq11}, n=2,3,4,5
\end{eqnarray} 

\noindent where $x=rcos\phi$ and $y=rsin\phi$. Eq.\ref{eq11} also determine the participant plane angle $\psi_n$. Asymmetry measures,
$\epsilon_2$ and $\epsilon_3$ are called eccentricity and triangularity. $\epsilon_4$ and $\epsilon_5$ essentially measures the squareness and five-sidedness of the initial distribution. In the following,
$\epsilon_4$   will be called rectangularity. In the same vein, $\epsilon_5$ will be called pentangularity. Fourth flow coefficient $v_4$ is generally referred as hexadecpolar flow. In following, we refer it as the rectangular flow, which is more appropriate. $v_5$ will be referred as the pentangular flow.

\section{Results}\label{sec4}

\subsection{Centrality dependence of charged particles multiplicity and $p_T$ spectra}

We have simulated 0-10\%, 10-20\%, 20-30\%, 30-40\% and 40-50\% Au+Au collisions at 
$\sqrt{s}_{NN}$=200 GeV. In each collision centrality, we have simulated 
 $N_{event}$=1000 events.  The constant $\epsilon_0$ is fixed to reproduce experimental charged particles multiplicity in 0-10\% collision. It was then kept fixed for all the other collision centralities. In Fig.\ref{F1}, simulated charged particles multiplicities are compared with the PHENIX data \cite{Adler:2004zn}. Once the model parameters are fixed to reproduce experimental multiplicity in 0-10\% collision, event-by-event simulations well reproduces the experimental multiplicity in other collision centralities.
We do note that in peripheral collisions, simulated multiplicity overestimate the experimental multiplicity by  $\sim$10-15\%. 

Even though charged particles multiplicities are well reproduced, the model simulations 
failed to reproduce charged particles $p_T$ spectra, in particular in the high $p_T$ region. In Fig.\ref{F2}, model simulations for charged particles $p_T$ spectra, in 0-10\%, 10-20\%, 20-30\%, 30-40\% and 40-50\% collision centralities are compared with the PHENIX measurements  \cite{Adler:2003au}. Simulated spectra explains the experimental data only up to $p_T\approx$1 GeV. In all the collision centralities, at higher $p_T$, model produces less particles than in experiment. The results
are interesting. It is well known that, compared to smooth hydrodynamics, in event-by-event hydrodynamics, $p_T$ spectrum is hardened \cite{arXiv:1104.0650}. Still the hardening is not enough to produce requisite number of particles at large $p_T$. 
It is also well known that $p_T$ spectra is hardened in viscous fluid. Better fit to 
charged particles $p_T$ spectra at large $p_T$ is expected if viscous rather than ideal fluid is formed in the collisions.
We do note that in the present simulations, we have
not made any conscientious attempt to fit the $p_T$, spectra. Only the
  charged particles multiplicity in 0-10\% collision was fitted. Varying other parameters e.g. initial time, initial fluid velocity, freeze-out temperature etc. fit to charged particles $p_T$ spectra may be improved. 
  
  \begin{figure}[t]
\center
\resizebox{0.4\textwidth}{!}{%
\includegraphics{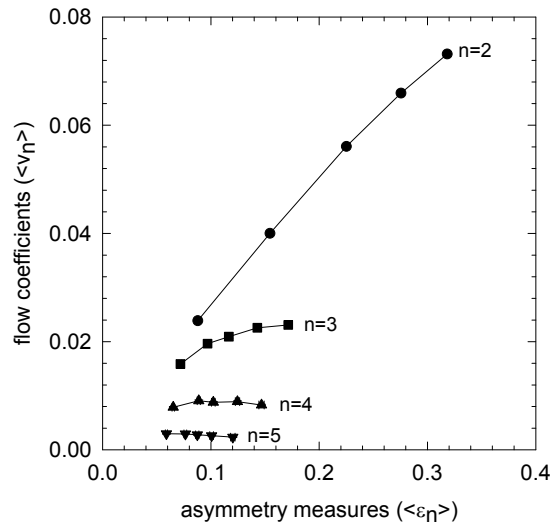}
}
\caption{Event averaged flow coefficients ($v_n$) against the   asymmetry measures,
($\epsilon_n$) for n=2-5. }
\label{F4}
\end{figure} 

\subsection{Centrality dependence of flow coefficients}

\begin{figure}[t]
\center
\resizebox{0.40\textwidth}{!}{%
\includegraphics{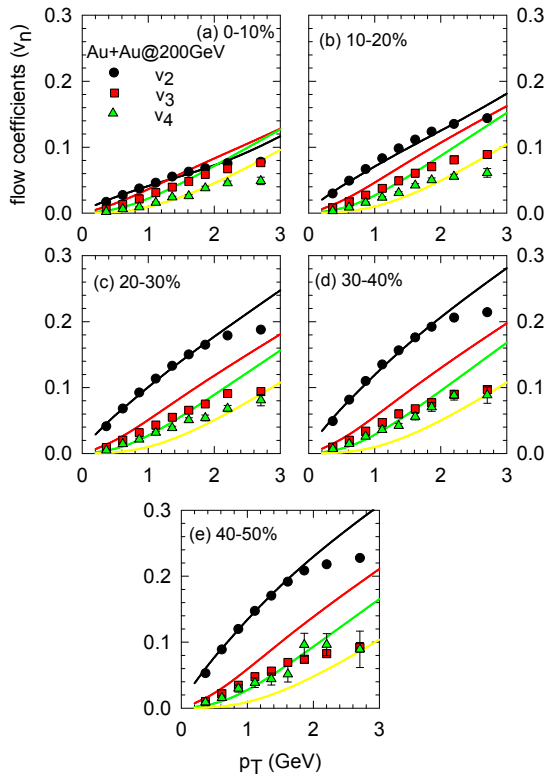} 
}
\caption{(color online) In five panels, event-by-event simulations for flow coefficients $v_n$, n=2-5 in Au+Au collisions are shown. The black, red, green and yellow lines are respectively for elliptic flow $v_2$, triangular flow $v_3$, rectangular flow $v_4$ and pentangular flow $v_5$. The black circles, red squares and green triangles are PHENIX measurements for elliptic, triangular and rectangular flow in Au+Au collisions at RHIC.}
\label{F5}
\end{figure} 

\subsubsection{Integrated flow coefficients}

Integrated flows are one of the important observables in heavy ion collisions. As discussed earlier, initial spatial symmetry is converted into momentum asymmetry, which is quantified in terms of different flow coefficients.  
For example, elliptic flow ($v_2$) is  response of an initial eccentricity ($\epsilon_2$) of the collision zone. Triangular flow ($v_3$) is response of initial triangularity ($\epsilon_3$) of the medium. Similarly, higher flow coefficients $v_4$ and $v_5$ are response of initial rectangularity ($\epsilon_4$) and pentangularity ($\epsilon_5$)of the initial medium.
In Fig.\ref{F3}, the black circles are PHOBOS measurements  \cite{Back:2004mh} for centrality dependence of elliptic flow ($v_2$). $v_2$ increases rapidly as the collisions become more and more peripheral. Present simulations for $v_2$ in event-by-event hydrodynamics are shown as red circles. The simulations results agree well with the experimental data. In Fig.\ref{F3},
simulation results for (integrated) triangular flow ($v_3$), rectangular flow ($v_4$) and pentangular
flow ($v_5$) are also shown. Triangular flow also increases as the collisions become more and more peripheral.  However, rate of increase is much slower than that for elliptic flow. $v_4$ and $v_5$ on the otherhand appears to be approximately independent of the collision centrality. From central 0-10\% to peripheral 40-50\% collisions, they change by less than a few percent. 

   \begin{figure}[t]
\center
\resizebox{0.3\textwidth}{!}{%
\includegraphics{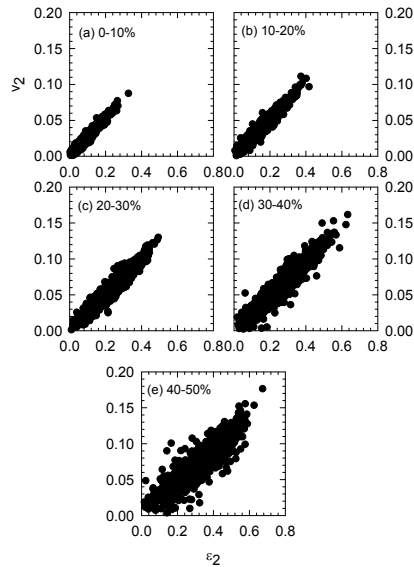} 
}
\caption{Correlation between elliptic flow ($v_2$) and initial eccentricity ($\epsilon_2$). Simulation results for $v_2$ is plotted against the initial eccentricity for $N_{event}$=1000 events. For a perfect correlation $v_2 \propto \epsilon_2$, all the points should lie on a straight line.}
\label{F6}
\end{figure}

In smooth hydrodynamics, elliptic flow in Au+Au collisions has been investigated in detail. Approximately, elliptic flow is proportional to initial eccentricity $\epsilon_2$. Dependence of the event averaged flow coefficients ($\la v_n \ra$,n=2-5) on the asymmetry measures ($\la \epsilon_n \ra$)
in event-by-event hydrodynamics is shown in Fig.\ref{F4}. The symbols, from left to right corresponds to 0-10\%, 10-20\%, 20-30\%, 30-40\% and 40-50\% Au+Au collisions. As expected, asymmetry measures increases with collision centralities. The increase is most in $\epsilon_2$, by a factor of $\sim$3.5 from 0-10\% collision centrality to 40-50\% centrality. In other asymmetry measures, $\epsilon_n$, n=3-5, the increase is more modest, factor of $\sim$ 2-2.5 only. As it is in smoothed hydrodynamics, in event-by-event hydrodynamics also,
elliptic flow increase, approximately linearly, with the initial eccentricity,
$\la v_2 \ra \propto \la \epsilon_2 \ra$. Higher flow coefficients, $v_3$ also increase with initial triangularity, however, the increase evidently  is not linear.
Still higher flow coefficients $v_4(v_5) $, approximately remains the same in all the collision centralities (as already shown in Fig.\ref{F3}), they 
 appear to be independent of the asymmetry measures, $\epsilon_4(\epsilon_5)$.  
 Approximate centrality independence of   higher flow coefficients, $v_4$ and $v_5$ in event-by-event hydrodynamics indicate that unlike the elliptic or triangular flow, rectangular flow $v_4$ or pentangular flow $v_5$ may not be related to initial asymmetry measure of the collision zone. Later, we will discuss the issue in more detail.

\subsection{Differential flow coefficients}

Differential flow coefficients are very sensitive observables and a model is well tested by comparing its predictions against experimental differential   flow data. In Fig.\ref{F5}, in five panels (a)-(e), event-by-event hydrodynamic simulations for the differential flow coefficients, in 0-10\%, 10-20\%, 20-30\%, 30-40\% and 40-50\% Au+Au collisions  are shown. In each panel, the black, red, green and yellow lines are the simulation results for elliptic flow, triangular flow, rectangular flow and pentangular flow respectively. We have shown only the event averaged values. In each collision centralities,  PHENIX measurements  \cite{Adare:2010ux}\cite{Adare:2011tg}\cite{Lacey:2011av} for the elliptic, triangular and   rectangular flow are shown as the black circles, red squares and yellow triangles. Simulations do reproduce the trend of the data,
$v_2>v_3>v_4$. Event-by-event hydrodynamic simulations for the differential elliptic flow in Au+Au collisions agree well with the PHENIX data in all the collision centralities. In peripheral collisions, at $p_T>$ 2 GeV, experimental data are marginally over predicted. We have simulated Au+Au collisions in the   ideal fluid approximation. 
If instead of ideal fluid, viscous fluid is produced, better agreement with data is expected. Indeed, explicit event-by-event hydrodynamic simulations \cite{Qiu:2011hf}\cite{Schenke:2011zz}\cite{Chaudhuri:2011qm} do indicate that the event averaged flow coefficients reduces with viscosity.
In any case the agreement with data for elliptic flow measurements is much better for event-by-event ideal hydrodynamics compared to ideal hydrodynamic calculations with smooth CGC/Glauber model initial conditions  \cite{Roy:2012jb}\cite{Chaudhuri:2009hj}.

Even though simulation results for elliptic flow reasonably well agree with the PHENIX  experiment, simulation results for triangular ($v_3$) and rectangular flow ($v_4$) appear to over predict the PHENIX data for the same. Interestingly, triangular flow is more over predicted than the rectangular flow. Also, the discrepancy between simulations and experiment is more in peripheral collisions than in central collisions. For example, in 0-10\% collision, simulated $v_3$,
in the $p_T$ range 1-2 GeV, over predict the PHENIX data by $\sim$ 30\%. In  30-40\% collision, the data are over predicted by $\sim$60\% or more. In 30-40\% collisions, rectangular flow, in the $p_T$ range 1-2 GeV, is overpredicted by 5-10\% only. Here again, better agreement with data is expected if instead of ideal fluid, viscous fluid is formed in Au+Au collisions.

  \begin{figure}[t]
\center
\resizebox{0.3\textwidth}{!}{%
\includegraphics{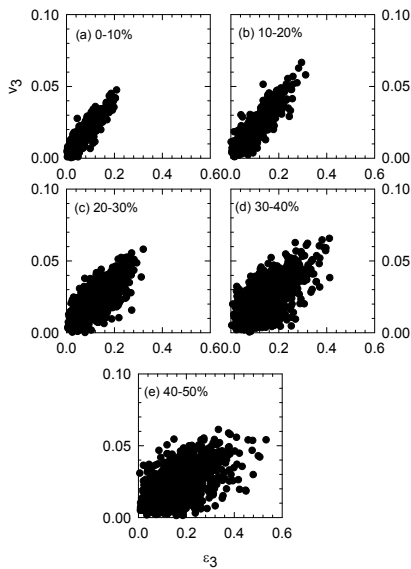} 
}
\vspace{0.4cm} 
\caption{Same as in Fig.\ref{F6} but for triangular flow ($v_3$) and initial triangularity ($\epsilon_3$).}
\label{F7}
\end{figure}

In Fig.\ref{F5}, the yellow lines represents the pentangular flow $v_5$. We did not find any experimental data for the pentangular flow. Following the trend of the simulation results for higher harmonic $v_3$, $v_4$, which are overpredicted in simulations, we do expect that the present simulation also over predict $v_5$. Present simulations then suggest that in experiments, in a peripheral 40-50\% Au+Au collisions, in the $p_T$ range 1-2 GeV,   $\sim$2-5\% or less pentangular flow may be expected.

\subsection{Correlation between (integrated) flow coefficients and initial asymmetry measures}  
 
 Recently, in    \cite{Chaudhuri:2011pa}\cite{Chaudhuri:2012wn}, correlation between integrated flow coefficients ($v_n$) and initial asymmetry measures ($\epsilon_n$) of the collision zone was studied in event-by-event hydrodynamics. It was shown that while elliptic flow remain strongly correlated with initial eccentricity, correlations  between the higher flow coefficients $v_n$ and initial asymmetry measures $\epsilon_n$, n=3,4,5, are much more weak. In \cite{Chaudhuri:2011pa}\cite{Chaudhuri:2012wn} correlations between flow coefficients and asymmetry measures, in  a single collision centrality, were studied. How the correlations are affected, as a function of collision centralities were not studied. 

  \begin{figure}[t]
\center
\resizebox{0.3\textwidth}{!}{%
\includegraphics{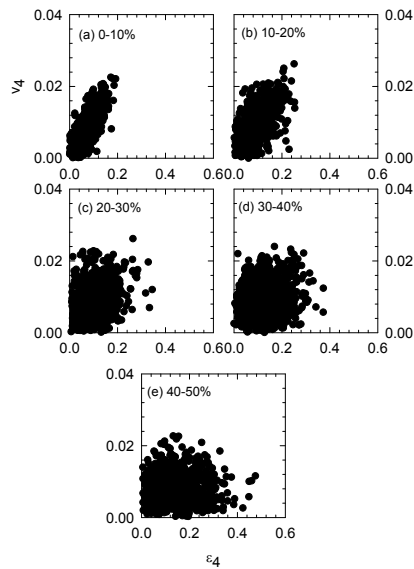} 
}
\caption{Same as in Fig.\ref{F6} but for rectangular flow ($v_4$) and initial rectangularity ($\epsilon_4$).}
\label{F8}
\end{figure}   

\begin{figure}[t]
\center
\resizebox{0.3\textwidth}{!}{%
\includegraphics{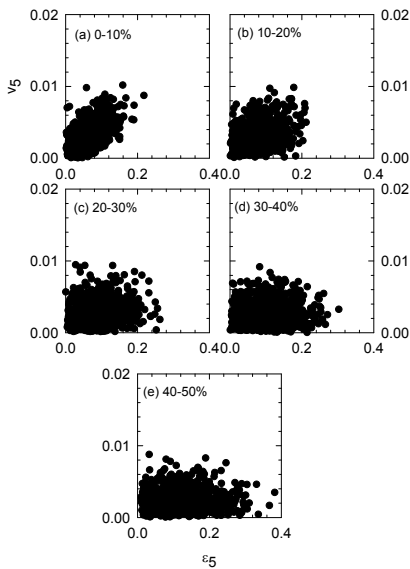} 
}
\caption{Same as in Fig.\ref{F6} but for pentangular flow ($v_5$) and initial pentangularity ($\epsilon_5$).}
\label{F9}
\end{figure}

In Fig.\ref{F6}, we have plotted the simulated elliptic flow ($v_2$) against the initial eccentricity ($\epsilon_2$) in $N_{event}$=1000 events. If $v_2$ is perfectly correlated with $\epsilon_2$, all the points should lie on a straight line. One observes that in central collisions, elliptic flow is strongly correlated with eccentricity. The correlation is marginally reduced in more peripheral collisions. One can conclude that in event-by-event hydrodynamics also, the correlation between elliptic flow and initial eccentricity is strong, irrespective   of the collision centrality.
In Fig.\ref{F7}-\ref{F9},
results obtained for higher flow harmonics are shown. In central collisions, correlation between triangular flow ($v_3$) and initial triangularity ($\epsilon_3$) is strong, though degree of correlation appear to be less than that in elliptic flow. In more peripheral collisions however, correlation is significantly reduced. Correlation between  higher flow harmonics, $v_4(v_5)$ and initial asymmetry measure $\epsilon_4(\epsilon_5)$ even in central collisions is visibly much weaker  than the corresponding correlation between elliptic flow and initial eccentricity. The correlations deteriorate as the collisions become more and more peripheral. Indeed, from the scatter plot of $v_4$ and $v_5$ in peripheral collisions, it is difficult to claim that the flow coefficients are correlated with the asymmetry measures.

In \cite{Chaudhuri:2011pa} a quantitative measure was defined to quantify the correlation between flow coefficients and initial spatial asymmetry measure. A modified form is used here to quantify the correlation.  
For a perfect correlation, $v_n\propto \epsilon_n$ and simulated flow coefficients will fall on a straight line. Dispersion of the
flow coefficients around the best fitted straight line then gives a
measure of the correlation. We thus define a correlation measure
function $C_{measure}$,
  
\begin{equation}
C_{measure}(v_n)=1-\frac{\sum_i [ v_n^i(\epsilon_n) -v_{n,st.line}(\epsilon_n) ]^2}{\sum_i [ v^i_{random}(\epsilon) -v_{st.line}(\epsilon) ]^2}
\end{equation}

\begin{figure}[t]
\center
\resizebox{0.35\textwidth}{!}{%
\includegraphics{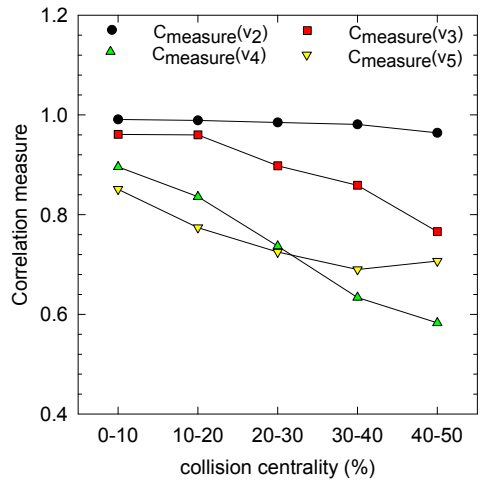} 
}
\caption{(color online) collision centrality dependence of the correlation measure (see text) for $v_2$, $v_3$, $v_4$ and $v_5$ }
\label{F10}
\end{figure}

$C_{measure}$ 
  essentially measures the dispersion of the simulated flow coefficients from the best fitted straight line, relative to completely random flow coefficients. It varies between 0 and 1. 
If flow coefficients are perfectly correlated then $v_n \propto \epsilon_n$ and $C_{measure}$ is identically unity. For completely random flow coefficients, $C_{measure}$=0. To obtain an even ground for comparison of $C_{measure}$ for different flow coefficients, the flow coefficients ($v_n$) and the asymmetry parameters ($\epsilon_n$) are scaled to vary between 0 and 1.   In Fig.\ref{F10}, we have shown the correlation measures for the flow coefficients as a function of collision centrality.  
The elliptic flow remain strongly correlated with initial eccentricity ($c_{measure}(v_2)\approx 0.95-0.99$) in 0-50\% collisions. In central, 0-10\%, 10-20\% collisions, triangular flow ($v_3$) is also strongly correlated with initial triangularity ($\epsilon_3$), ($c_{measure}(v_2)\approx 0.95$). Correlation is significantly reduced in more peripheral collisions and in 40-50\% collisions,
$c_{measure}(v_2)\approx 0.75$. In higher flow coefficients, correlation is even less
in peripheral collisions.

If departure of $C_{measure}$ from unity is interpreted as a measure of flow uncorrelated with the initial
asymmetry measure, for elliptic flow $v_2$, in 0-50\% collisions, less than $\sim$ 5\% of the flow is uncorrelated with initial eccentricity.  In higher flow coefficients, $v_n$, n=3-5, uncorrelated flow grow with collision centrality. For example, in rectangular flow $v_4$, uncorrelated flow grows from $\sim$10\% in 0-10\% collisions to $\sim$40\% in 40-50\% collision.

As it was discussed previously, present analysis of Au+Au data   also indicate that better description to the data will be obtained if viscous fluid rather than ideal fluid is formed in Au+Au collision. In   \cite{Chaudhuri:2011pa} viscous effects on the correlation between elliptic flow and initial eccentricity and between triangular flow and initial triangularity were studied. It was shown that the correlations reduce significantly in viscous fluid. In more realistic event-by-event hydrodynamic simulation of Au+Au collisions, with viscous fluid, correlation between higher flow coefficients and asymmetry measures then expected to reduce even more than obtained presently.

Compartively low $C_{measure}$ or equivalently, large uncorrelated higher flow harmonics is difficult to understand.
In ideal hydrodynamics,   final flow coefficients are related to the initial transverse energy density, or more appropriately on the pressure gradients only. We have even assumed zero initial velocity. Yet, the flow coefficients, $v_3$, $v_4$ and  $v_5$ in peripheral collisions are largely unrelated to the initial asymmetry measures. Elliptic flow on the other hand is perfectly correlated with initial asymmetry measure. Source of the uncorrelated flows in higher harmonics can not be discerned presently. Possibly, other aspects of the
initial density (higher moments or products of moments) are important in development of higher harmonics.  

\section{Summary and conclusions}\label{sec5}

To summarize, in event-by-event hydrodynamics, we have studied the centrality dependence of charged particles multiplicity, $p_T$ spectra, and flow coefficients (integrated and differential) in $\sqrt{s}_{NN}$=200 GeV Au+Au collisions. The initial energy density distributions are obtained from Monte-Carlo Glauber model. The Monte-Carlo Glauber model participant positions are smoothed with a Gaussian distribution of width $\sigma$=0.5 fm and normalized to reproduce experimental charged particles multiplicity in 0-10\% collision. We have simulated a large number of events, $N_{event}$=1000 in each collision centrality. Once the initial transverse energy density is fixed to reproduce multiplicity in 0-10\% collision, the model reproduces the experimental multiplicity in other collision centralities within reasonable accuracy.  
Experimental charged particles transverse momentum  spectra, however are reproduced in the model, only in a limited $p_T$ range, $p_T \leq $1GeV. At $p_T >$ 1 GeV, simulated spectra under predict the experiment. In the simulations, dissipative effects are neglected. Dissipative effect like (shear) viscosity, will enhances particle production, more at high $p_T$ than at low $p_T$. Better description to the $p_T$ spectra is expected if instead of ideal fluid, viscous fluid is produced in Au+Au collisions. 
We have also compared the model simulations for integrated and differential  flow coefficients with experimental data. Experimental (integrated) elliptic flow in Au+Au collisions are correctly reproduced in simulations.  The model also reasonably well reproduces the experimental differential elliptic flow in 0-10\%-40-50\% collisions.
In peripheral collisions, elliptic flow data however is overpredicted at high $p_T$. Higher flow coefficients $v_3$ and $v_4$ however are over predicted, more in peripheral than in central collisions. Here again, better description to the data is expected  if instead of ideal fluid, viscous fluid is produced in Au+Au collisions.

We have also studied correlation between (integrated) flow coefficients and initial asymmetry measures of the collision zone. In all the collision centralities (0-10\% to 40-50\%) elliptic flow is strongly correlated with 
the initial asymmetry measure, the eccentricity of the collision zone. The higher flow
coefficients however show much less correlation with the corresponding asymmetry measures. We have quantified the correlation and observe that with the exception for
elliptic flow, which remain strongly correlated in all the collision centralities, for the higher flow coefficients, $v_3$, $v_4$ and $v_5$, correlation reduces significantly in more peripheral collisions. It appears that in higher flow coefficients, a significant part of the flow is unrelated to the  initial asymmetry measures. 
The reason for the 
flow unrelated to the  initial asymmetry can not be discerned presently. One can only conclude that apart from the initial density distribution of collision zone, other aspects, e.g. higher moments or product of higher moments are also important for the development of higher harmonics.

\section*{Acknowledgments} 
RH,  VR and BM are supported by DAE-BRNS project Sanction No. 2010/21/15-BRNS/2026.

\end{document}